\documentclass[11pt]{article}
\usepackage{moriond,epsfig}
\usepackage{overpic}

\def\be{\begin{equation}}
\def\ee{\end{equation}}
\def\bea{\begin{eqnarray}}
\def\eea{\end{eqnarray}}

\newcommand{\mpt}{\ensuremath{/\!\!\!\!{p}_{\rm T}}}

\newcommand{\fb}{\ensuremath{{\rm fb}^{-1}}}

\newcommand{\pt}{\ensuremath{p_{\rm T}}}

\newcommand{\etal}{{\em et. al.}}

\newcommand{\GeV}{\ensuremath{\textnormal{GeV}}}
\newcommand{\TeV}{\ensuremath{\textnormal{TeV}}}

\newcommand{\dzero}{D\O}

\newcommand{\ttbar}{\ensuremath{t\bar{t}}}

\newcommand{\etmissx}{\ensuremath{E \kern-0.6em\slash_{\rm x}}}
\newcommand{\etmissy}{\ensuremath{E \kern-0.6em\slash_{\rm y}}}

\newcommand{\dm}{\ensuremath{\Delta m}}
\newcommand{\mtop}{\ensuremath{m_{\rm top}}}
\newcommand{\mt}{\ensuremath{m_t}}
\newcommand{\mtb}{\ensuremath{m_{\bar t}}}
\newcommand{\ejets}{\ensuremath{e+{\rm jets}}}
\newcommand{\mujets}{\ensuremath{\mu+{\rm jets}}}
\newcommand{\ljets}{\ensuremath{\ell+{\rm jets}}}

\newcommand{\sigtt}{\ensuremath{\sigma_{t\bar t}}}
\newcommand{\mpole}{\ensuremath{m_{\rm top}^{\rm pole}}}
\newcommand{\mmsbar}{\ensuremath{m_{\rm top}^{\overline{\rm MS}}}}
\newcommand{\mmc}{\ensuremath{m_{\rm top}^{\rm MC}}}
\newcommand{\prob}{\ensuremath{\mathcal{P}}}
\newcommand{\psig}{\ensuremath{\mathcal{P}_{\rm sig}}}
\newcommand{\kjes}{\ensuremath{k_{\rm JES}}}

\begin{document}
\vspace*{4cm}
\title{MEASUREMENTS OF THE TOP QUARK MASS AT THE TEVATRON}

\author{{\sc O. Brandt} on behalf of the {\sc CDF} and {\sc D0 Collaborations}}

\address{II. Physikalisches Institut, Friedrich-Hund-Platz 1,\\
G\"ottingen, Germany}

\maketitle\abstracts{
The mass of the top quark (\mtop) is a fundamental parameter of the standard model (SM). Currently, its most precise measurements are performed by the CDF and D0 collaborations at the Fermilab Tevatron $p\bar p$ collider at a centre-of-mass energy of $\sqrt s=1.96~\TeV$. We review the most recent of those measurements, performed on data samples of up to 8.7~\fb\ of integrated luminosity. The Tevatron combination using up to 5.8 fb$^{-1}$ of data results in a preliminary world average top quark mass of $m_{\rm top} = 173.2 \pm 0.9$~GeV. This corresponds to a relative precision of about 0.54\%. We conclude with an outlook of anticipated precision the final measurement of \mtop\ at the Tevatron.
\vspace{2mm}\\
PACS {\tt 14.65.Ha} -- Top quarks.
}

\section{Introduction}
The pair-production of the top quark was discovered in 1995 by the CDF and D0 experiments~\cite{bib:topdiscovery} at the Fermilab Tevatron proton-antiproton collider. Observation of the electroweak production of single top quarks was presented only two years ago~\cite{bib:singletop}. The large top quark mass and the resulting Yukawa coupling of almost unity indicates that the top quark could play a crucial role in electroweak symmetry breaking. Precise measurements of the properties of the top quark provide a crucial test of the consistency of the SM and could hint at physics beyond the SM. 

\begin{figure}
\centering
\begin{overpic}[height=0.335\textwidth]{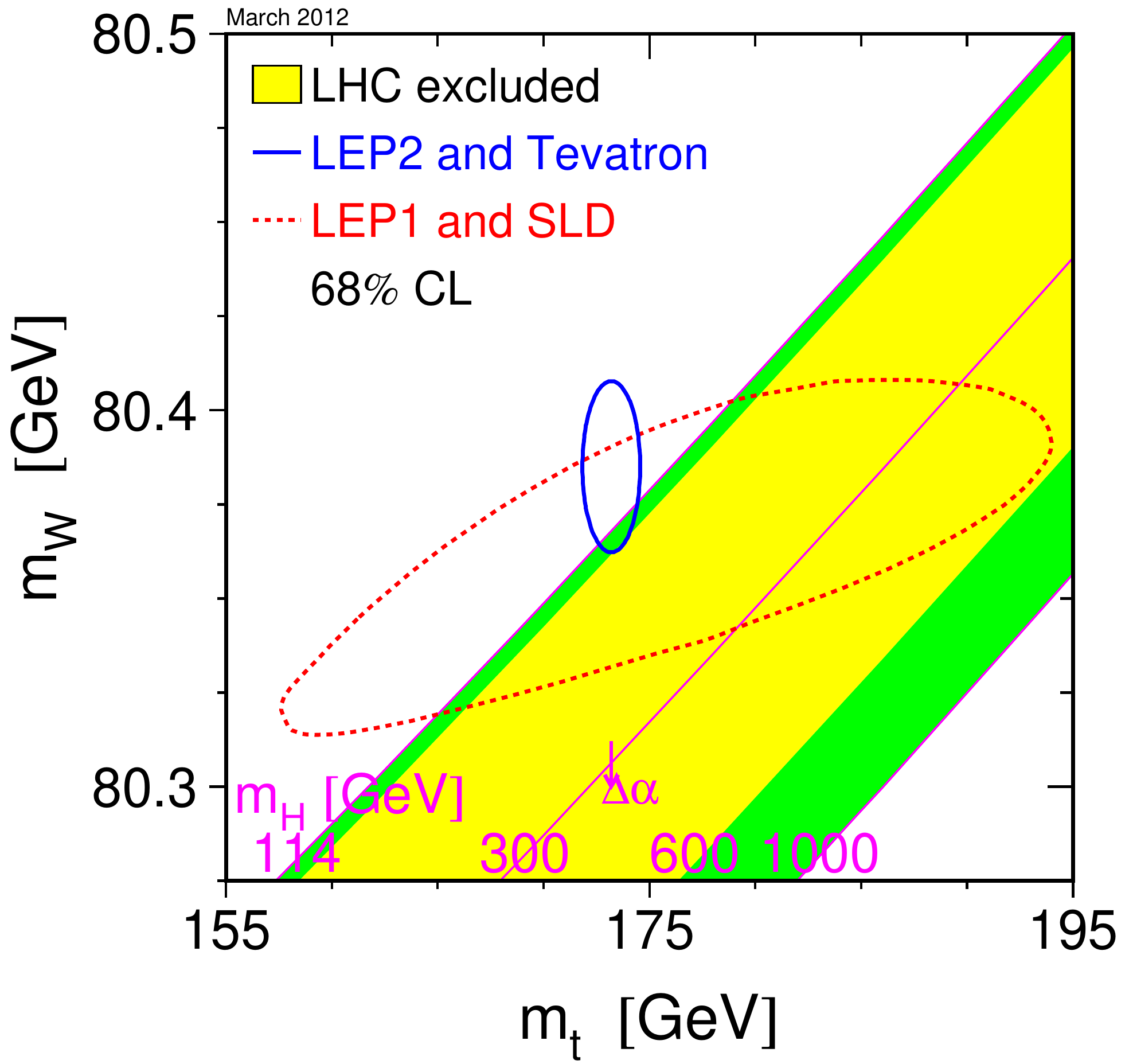}
\put(-1,2){(a)}
\end{overpic}
\qquad
\begin{overpic}[height=0.33\textwidth]{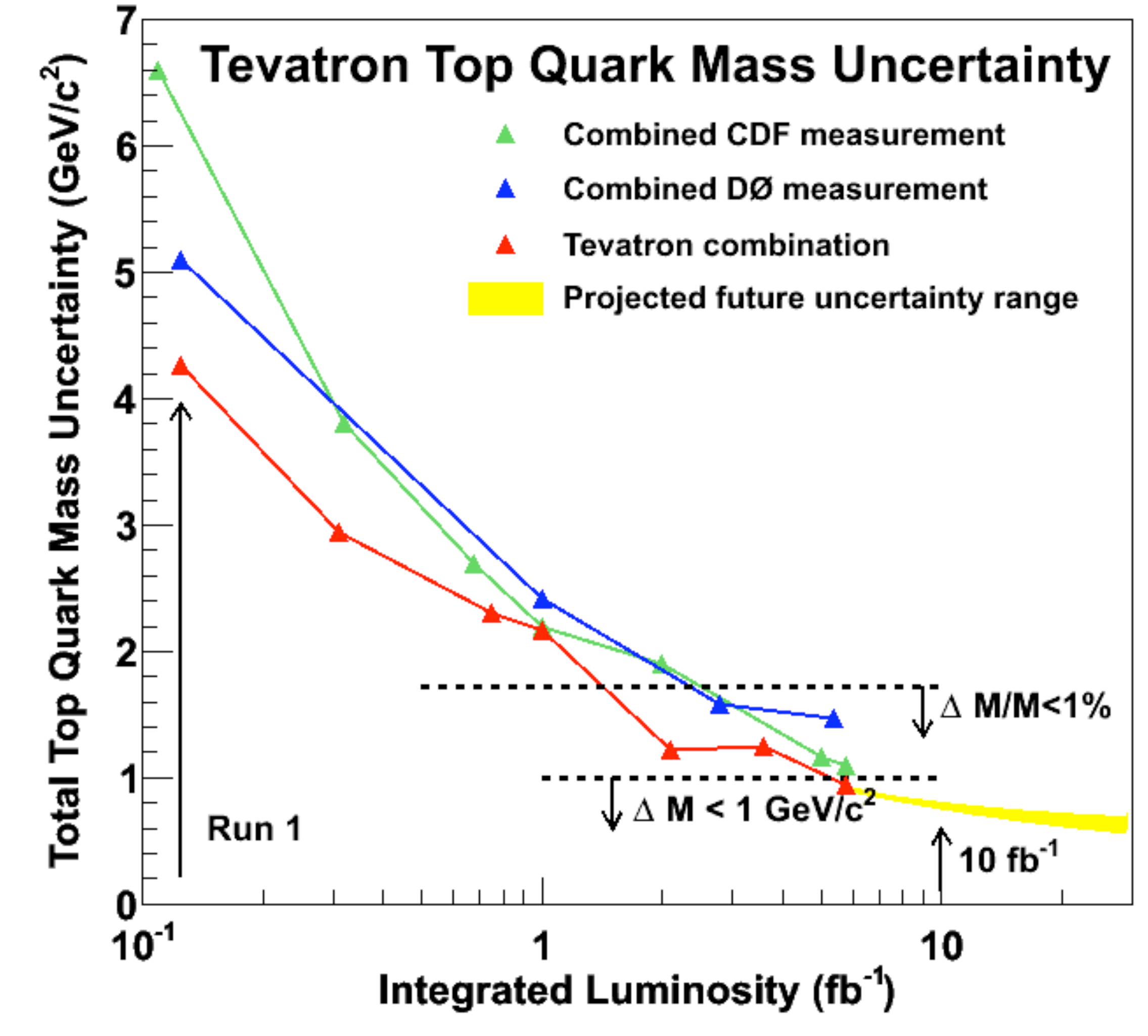}
\put(-1,2){(b)}
\end{overpic}
\caption{
\label{fig:mhiggs}
{\bf(a)} The constraint on mass of the SM Higgs boson from direct $\mtop$ and $m_W$ measurements in the $\mtop$,$m_W$ plane~$^9$. The red ellipsis indicates the 68\% CL contour. {\bf(b)} The anticipated precision on \mtop\ measurements at D0 and the Tevatron combination versus integrated luminosity.
}
\end{figure}

In the following, we review measurements of the top quark mass at the Tevatron, which is a fundamental parameter of the SM. Its precise knowledge, together with the mass of the $W$~boson ($m_W$), provides an important constraint on the mass of the postulated SM Higgs boson. This is illustrated in the \mtop,$m_W$ plane in Fig.~\ref{fig:mhiggs}, which includes the recent, most precise measurements of $m_W$~\cite{bib:wmasstalk}. A detailed review of measurements of the top quark mass is provided in Ref.~\cite{bib:reviewmtop}. Recent measurements of properties of the top quark other than \mtop\ at the Tevatron are reviewed in Ref.~\cite{bib:proptalk}. The full listing of top quark measurements at the Tevatron is available at public web pages~\cite{bib:toprescdf,bib:topresd0}.

At the Tevatron, top quarks are mostly produced in pairs via the strong interaction. 
By the end of Tevatron operation, about 10 fb$^{-1}$ of integrated luminosity per experiment were recorded by CDF and D\O, which corresponds to about 80k produced $\ttbar$ pairs. In the framework of the SM, the top quark decays to a $W$~boson and a $b$~quark nearly 100\% of the time, resulting in a $W^+W^-b\bar b$ final state from top quark pair production. 
Thus, $\ttbar$ events are classified according to the $W$ boson decay channels as ``dileptonic'', ``all--jets'', or ``lepton+jets''. More details on the channels and their experimental challenges can be found in Ref.~\cite{bib:xsec}, while the electroweak production of single top quarks is reviewed in Ref.~\cite{bib:singletoptalk}.

\section{Direct measurements of the top quark mass in \ljets\ final states}
D0's most precise measurement of \mtop\ is performed in $\ell+4{\,\rm jets}$ final state using the so-called {\em matrix element} (ME) method in 3.6~\fb\ of data~\cite{bib:mtoplj_d0}. This technique was pioneered by \dzero\ in Run~I of the Tevatron~\cite{bib:topmassd0nature}, and it calculates the probability that a given event, characterised by a set of measured observables $x$, comes from the \ttbar\ production given an \mtop\ hypothesis, or from a background process: 
$\prob_{\rm evt}(x)\propto f\prob_{\rm sig}(x,\mtop)+(1-f)\prob_{\rm bgr}$.
The dependence on $\mtop$ is explicitly introduced by calculating $\psig$ using the differential cross section ${\rm d}\sigma(y,\mtop)\propto|\mathcal{M}_{\ttbar}|^2(\mtop)$, where $\mathcal{M}_{\ttbar}$ is the leading order (LO) matrix element for \ttbar\ production:
\[
\psig(x,\mtop,\kjes) = \frac1{\sigtt^{\rm observed}}\cdot\int W(x,y,\kjes)~{\rm d}\sigma(y,\mtop)\,.
\]
Since ${\rm d}\sigma(y,\mtop)$ is defined for a set of parton-level observables $y$, the transfer function $W(x,y,\kjes)$ is used to map them to the reconstruction-level set $x$. This accounts for detector resolutions and acceptance cuts, and introduces explicitly the dependence on the jet energy scale (JES) via an overall scaling factor \kjes. The uncertainty on the JES, which is almost fully correlated with \mtop, is around 2\% or larger. Therefore, an {\em in situ} calibration is performed by requiring that the mass of the dijet system assigned to the parton pair from the hadronically decaying $W$ boson be $m_{jj}=80.4~\GeV$. Thus, $\mtop$ and $\kjes$ are extracted simultaneously. This reduces the uncertainty from the JES to about 0.5\%, decreasing with the number of selected $\ttbar$ events. The measurement is performed in events with four jets, resulting in 24 possible jet-parton assignments. All 24 assignments are summed over, weighted according to the consistency of a given assignment with the $b$-tagging information. $\prob_{\rm bgr}$ is calculated using the VECBOS matrix element for $W+4$~jets production. Generally, the ME technique offers a superior statistical sensitivity as it uses the full topological and kinematic information in the event in form of 4-vectors. The drawback of this method is the high computational demand.

D0 measures $\mtop=174.9 \pm 0.8~({\rm stat}) \pm 0.8~({\rm JES}) \pm 1.0~({\rm syst})~\GeV$, corresponding to a relative uncertainty of 0.9\%. The dominant systematic uncertainties are from modeling of underlying event activity and hadronisation, as well as the colour reconnection effects. On the detector modeling side, diffential uncertainties on the JES which are compatible with the overall $\kjes$ value from {\em in situ} calibration, and the difference between the JES for light and b-quark jets are dominant. This picture is representative for all \mtop\ measurements in \ljets\ final states shown here.

CDF employs the ME technique similar to that used at D0 to measure \mtop\ on a dataset corresponding to 5.6~\fb and finds $\mtop=173.0 \pm 0.7~({\rm stat}) \pm 0.6~({\rm JES}) \pm 0.9~({\rm syst})~\GeV$~\cite{bib:mtoplj_cdf}. Most notable differences from the D0 measurement are: (i)~background events present in the data sample are accounted for on {\em average} rather than on an event-by-event basis using a likelihood based on a neural network output, (ii)~the contribution of ``mismeasured'' signal events, where one of the jets cannot be matched to a parton, is reduced with a cut on the aforementioned likelihood.

Currently, the world's best single measurement of \mtop\ is performed by CDF in \ljets\ final states using the so-called {\em template} method to analyse the full dataset of 8.7~\fb~\cite{bib:mtopljtempl_cdf}. The basic idea of the template method is to construct ``templates'', i.e.\ distributions in a set of variables~$x$, which are sensitive to \mtop, for different mass hypotheses, and extract \mtop\ by matching them to the distribution found in data, e.g.\ via a maximum likelihood fit. CDF minimises a $\chi^2$-like function to kinematically reconstruct the event for jet-parton assignments consistent with the $b$-tagging information. To extract \mtop\ and calibrate the JES {\em in-situ}, three-dimensional templates are defined in the fitted \mtop\ of the best jet-parton assignment,
the fitted \mtop\ of the second-best assignment, and
the fitted invariant mass of the dijet system from the hadronically decaying $W$ boson. CDF finds $\mtop=172.9~ \pm 0.7~({\rm stat}) \pm 0.8~({\rm syst})~\GeV$.

\section{Direct measurement of the top quark mass in all-hadronic final states}
The third most statistically significant contribution to the current Tevatron average of \mtop\ comes from a measurement in $6\leq N_{\rm jets}\leq8$ final states by CDF using 5.8\,\fb\ of data~\cite{bib:mtophad_cdf}. The main challenge is the high level of the background contribution from QCD multijet production: the $S:B$ ratio is about $1:1200$ after a multijet trigger requirement. Therefore, a discrimination variable $\mathcal{D}_{\rm NN}$ is constructed with a multilayered neural network (NN). Beyond typical kinematic and topological variables, also jet shape variables  which provide discrimination between quark and gluon jets, are used as inputs. To enhance the purity of the sample and to reduce the number of combinatoric possibilities, $b$ tagging is applied. For each jet--parton assignment, a $\chi^2$ is constructed which accounts for: the consistency of the two dijet pairs with the reconstructed $m_W$, the consistency of the $jjb$ combinations with the reconstructed \mtop, and the consistency of the individual fitted jet momenta with the measured ones, within experimental resolutions. The final sample for top mass extraction is defined by $\mathcal{D}_{\rm NN}>0.97~(0.84)$ for events with 1~($\geq2$) $b$~tags, yielding a signal to background ratio of $1:3~(1:1)$. The measured value is $\mtop=172.5~ \pm 1.4~({\rm stat}) \pm 1.4~({\rm syst})~\GeV$. Beyond systematic uncertainties relevant in \ljets\ final states, potential biases from the data-driven background model pose a notable contribution to the total uncertainty.

\section{Direct measurement of the top quark mass in dilepton final states}
The world's most precise measurement of \mtop\ in dilepton final states is performed by D0 using 5.4~\fb\ of data~\cite{bib:mtopll_d0}. Leaving \mtop\ as a free parameter, dilepton final states are kinematically underconstrained by one degree of freedom, and the so-called neutrino weighting algorithm is applied for kinematic reconstruction. It postulates distributions in rapidities of the neutrino and the antineutrino, and calculates a weight, which depends on the consistency of the reconstructed $\vec\pt^{\,\nu\bar\nu}\equiv\vec\pt^{\,\nu}+\vec\pt^{\,\bar\nu}$ with the measured missing transverse momentum $\mpt$ vector, versus \mtop. D0 uses the first and second moment of this weight distribution to define templates and extract \mtop. To reduce the systematic uncertainty, the {\em in situ} JES calibration in \ljets\ final states~\cite{bib:mtoplj_d0} is applied, accounting for differences in jet multiplicity, luminosity, and detector ageing. After calibration and all corrections, $\mtop=174.0~ \pm 2.4~({\rm stat}) \pm 1.4~({\rm syst})~\GeV$ is found.

\section{Measurement of \mtop\ from the \ttbar\ production cross-section}
The $\ttbar$ production cross section (\sigtt) is correlated to \mtop. This can be used to extract \mtop\ by comparing the measured \sigtt\ to the most complete to--date, fully inclusive theoretical predictions, assuming the validity of the SM. Such calcualtions offer the advantage of using mass definitions in well-defined renormalisation schemes like $\mmsbar$ or $\mpole$. In contrast, the main methods using kinematic fits utilise the mass definition in MC generators $\mmc$, which cannot be translated into $\mmsbar$ or $\mpole$ in a straightforward way.
D0 uses 5.3~\fb\ of data to measure $\sigtt$ and extracts \mtop~\cite{bib:mtoppole_d0} using theoretical calculations for $\sigtt$ like the next-to-leading order (NLO) calculation with next-to-leading logarithmic (NLL) terms resummed to all orders~\cite{bib:signlo}, an approximate NNLO calculation~\cite{bib:signnlo}, and others. For this, a correction is derived to account for the weak dependence of measured $\sigtt$ on $\mmc$. The results for \mpole are presented in Fig.~\ref{fig:mpole}, and can be summarised as follows: $\mpole=163.0^{+5.1}_{-4.6}~\GeV$ and $\mpole=167.5^{+5.2}_{-4.7}~\GeV$ for Ref.~\cite{bib:signlo} and~\cite{bib:signnlo}, respectively. 

\section{Measurements of the mass difference between the $t$ and $\bar t$ quarks}
\begin{figure}
\centering
\begin{overpic}[height=0.3\textwidth]{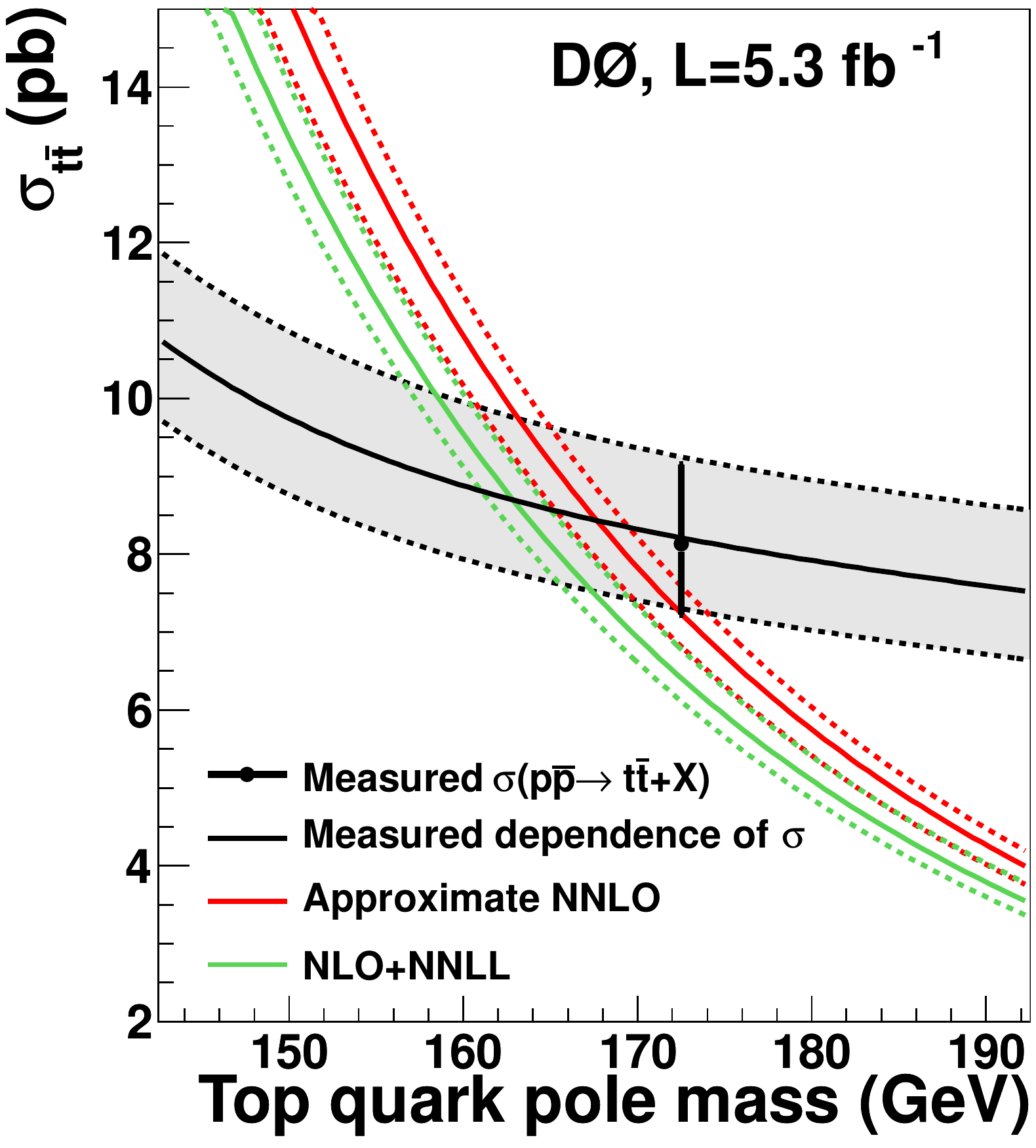}
\put(-1,2){(a)}
\end{overpic}
\qquad
\begin{overpic}[height=0.3\textwidth]{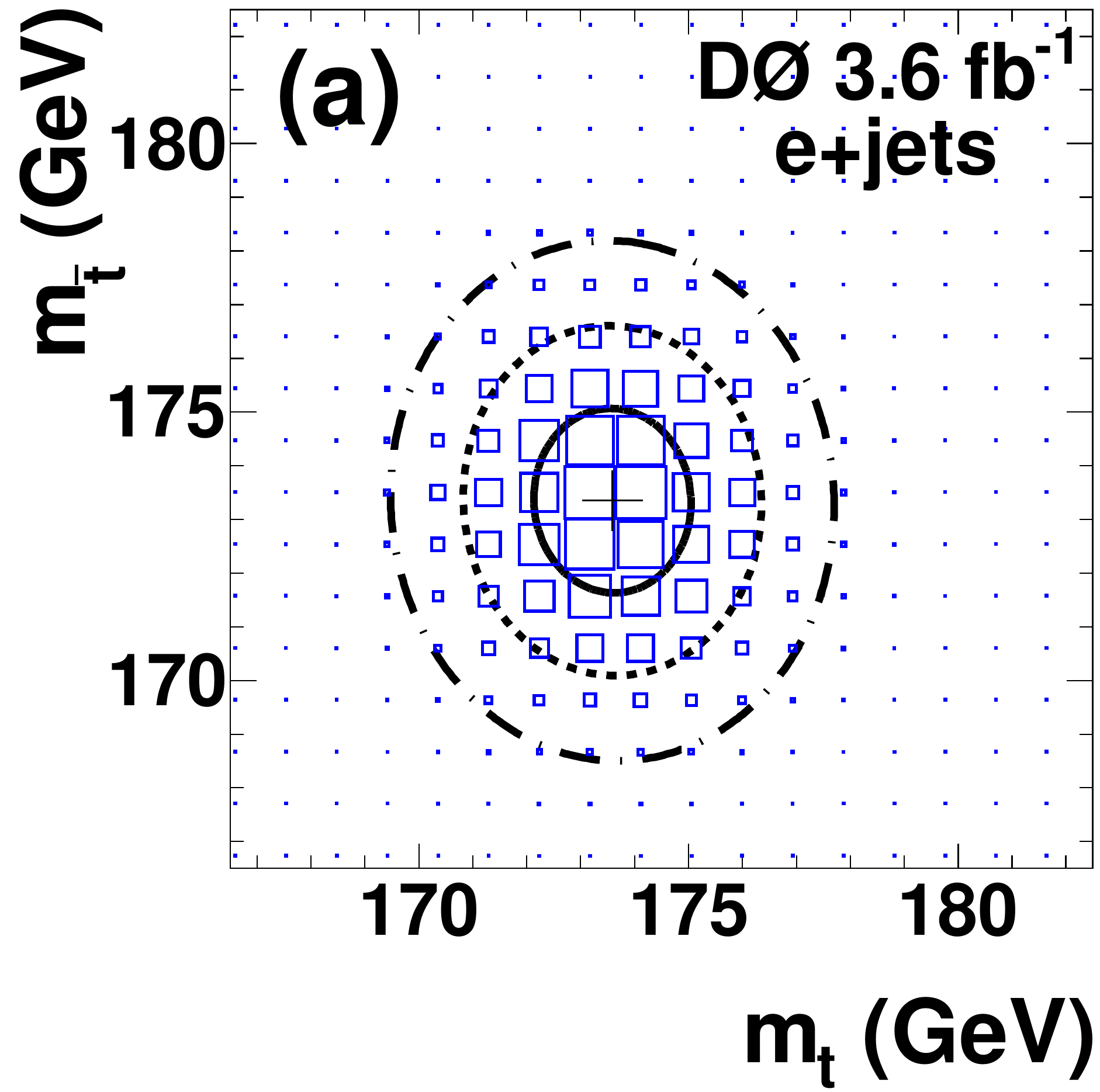}
\put(-1,2){(b)}
\end{overpic}
\qquad
\begin{overpic}[height=0.3\textwidth]{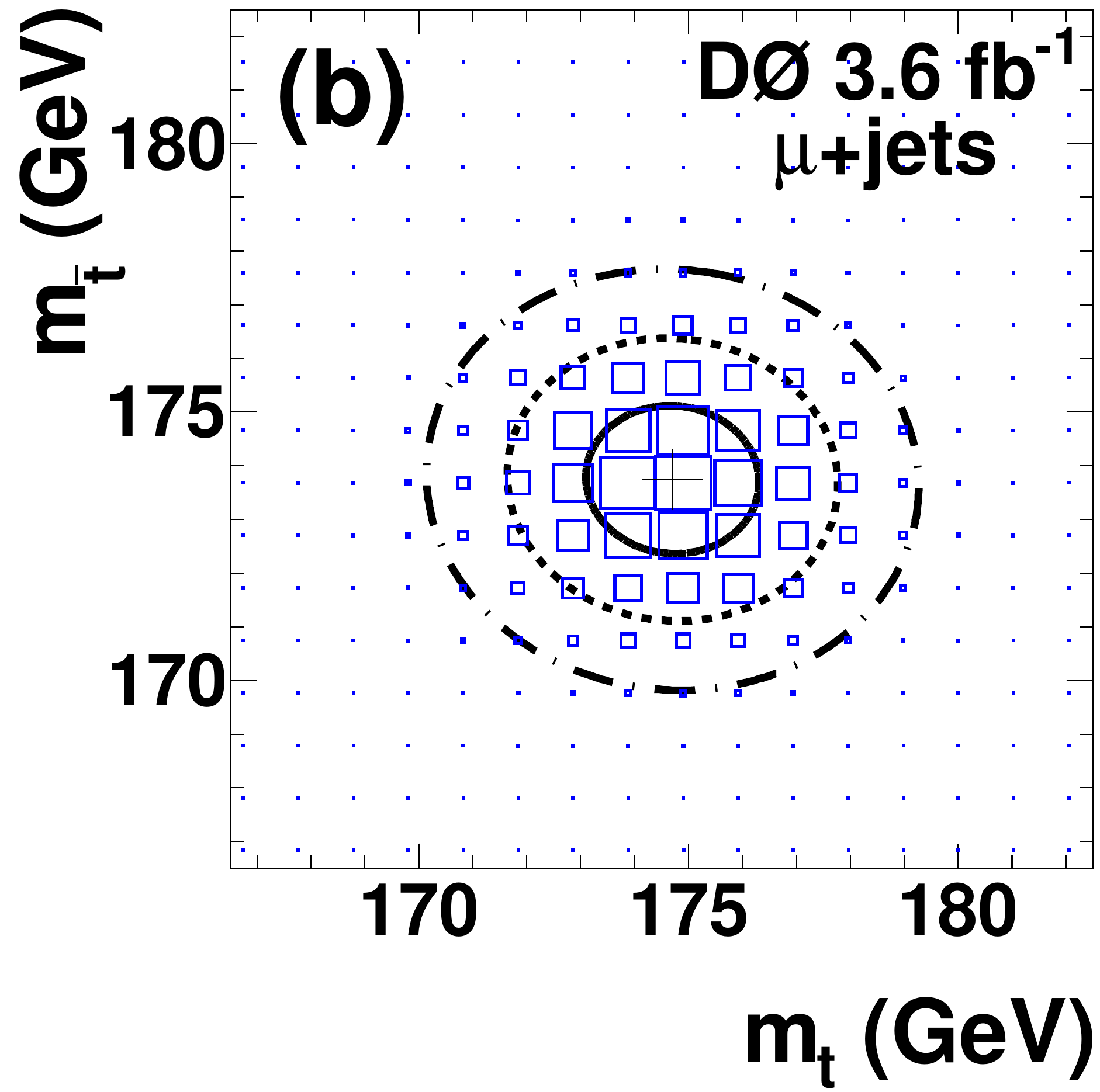}
\put(-1,2){(c)}
\end{overpic}
\caption{
\label{fig:mpole}
{\bf(a)} \sigtt\ measured by D0 using 5.3~\fb\ (black line) and theoretical NLO+NNLL~$^{17}$ (green solid line) and approximate NNLO~$^{18}$ (red solid line) predictions as a function of $\mpole$. The experimental acceptance correction assumes $\mmc=\mpole$. The gray band corresponds to the total uncertainty on measured $\sigtt$. The dashed lines indicate theoretical uncer\-tainties from the choice of scales and parton distribution functions. {\bf(b)} \mt\ and \mtb\ measured by D0 directly~and independently using 3.6~\fb in \ejets\ final states. The solid, dashed, and dash-dotted lines represent the 1, 2, and 3 SD contours. {\bf(c)} same as (b) but for \mujets.
}
\end{figure}

The invariance under $\mathcal{CPT}$ transformations is a fundamental property of the SM. $m_{\rm particle}\neq m_{\rm antiparticle}$ would constitute a violation of $\mathcal{CPT}$, and has been tested extensively in the charged lepton sector. Given its short decay time, the top quark offers a possiblity to test $\mt=\mtb$ at the percent level, which is unique in the quark sector. D0 applies the ME technique to measure $\mt$ and $\mtb$ directly and independently using 3.6~\fb\ of data, and finds $\dm\equiv\mt-\mtb=0.8\pm1.8~\GeV$~\cite{bib:dm_d0}, in agreement with the SM prediction. The results are illustrated in Fig.~\ref{fig:mpole}. With 0.5~\GeV, the systematic uncertainty on \dm\ is much smaller than that on $\mtop$ due to cancellations in the difference, and is dominated by the uncertainty on the difference in calorimeter response to $b$ and $\bar b$ quark jets. CDF uses a template-based method and a kinematic reconstruction similar to that in Ref.~\cite{bib:mtopljtempl_cdf} to measure $\dm$ directly given the constraint $\frac{\mt+\mtb}2\equiv172.5~\GeV$ from 8.7~\fb\ of data, and finds $\dm=-2.0\pm1.3~\GeV$~\cite{bib:dm_cdf}.

\vspace{-3mm}

\section{Tevatron combination and outlook}
Currently, the world's most precise measurements of \mtop\ are performed by CDF and D0 collaborations in \ljets\ final states. The preliminary Tevatron combination using up to 5.8 fb$^{-1}$ of data results in $m_{\rm top} = 173.2 \pm 0.9$~GeV~\cite{bib:combi}, corresponding to a relative uncertainty of 0.54\%.

With about 10.5~\fb\ recorded, the precision on \mtop\ is expected to further improve, especially at D0, where only 3.6~\fb\ are used in the flagship measurement in \ljets\ final states. This applies not only to the statistical uncertainty, but also to several systematic uncertainties due to the limited size of calibration samples, like e.g.\ some components of the JES. Moreover, efforts are underway to better understand systematic uncertainties from the modeling of \ttbar\ signal, in particular the dominating uncertainty from different hadronisation and underlying event models. We look forward to exciting updates of \mtop\ measurements presented here.

With uncertainties approaching $\mathcal{O}(\GeV)$ at the LHC~\cite{bib:mtoptalk}, we strongly advocate to start preparations towards the first world-wide combination of the measurements of the top quark mass including ATLAS and CMS results.

\section*{Acknowledgments}
I would like to thank my collaborators from the CDF and D0 experiments for their help in preparing this article. I also thank the staffs at Fermilab and collaborating institutions, as well as the CDF and D0 funding agencies.


\section*{References}
\begin{thebibliography}{99}

\bibitem{bib:topdiscovery}
F. Abe \etal\ (CDF Coll.), Phys. Rev. Lett. {\bf 74}, 2626 (1995), 
S.~Abachi~\etal\ (D0 Coll.), Phys. Rev. Lett. {\bf 74}, 2632 (1995).

\bibitem{bib:singletop}
T. Aaltonen \etal\ (CDF Coll.), Phys. Rev. Lett. {\bf 103}, 092001 (2009),
V.~M.~Abazov~\etal\ (D0 Coll.), Phys. Rev. Lett. {\bf 103}, 092002 (2009).

\bibitem{bib:wmasstalk}
R. Lopes de S\`a, these proceedings.

\bibitem{bib:reviewmtop}
A.~B. Galtieri~\etal, arXiv:1109.2163 [hep-ex] (2011).

\bibitem{bib:proptalk}
D. Mietlicki, these proceedings; A. Lister, these proceedings.

\bibitem{bib:toprescdf}
\verb|http://www-cdf.fnal.gov/physics/new/top/public_mass.html|

\bibitem{bib:topresd0}
\verb|http://www-d0.fnal.gov/Run2Physics/WWW/results/top.htm|,
\verb|http://www-d0.fnal.gov/Run2Physics/WWW/documents/Run2Results.htm|.

\bibitem{bib:xsec}
I. Aracena, these proceedings.

\bibitem{bib:singletoptalk}
B. Wu, these proceedings; R.~G. Suarez, these proceedings.

\bibitem{bib:ewwk}
\verb|http://lepewwg.web.cern.ch/LEPEWWG/|.

\bibitem{bib:mtoplj_d0}
V. M. Abazov \etal\ (D0 Coll.), Phys. Rev. D {\bf84}, 032004 (2011).

\bibitem{bib:topmassd0nature}
V. M. Abazov \etal\ (D0 Coll.),
Nature \textbf{429}, 638 (2004).

\bibitem{bib:mtoplj_cdf}
T. Aaltonen \etal\ (CDF Coll.), Phys. Rev. Lett. {\bf105}, 252001 (2010).

\bibitem{bib:mtopljtempl_cdf}
T. Aaltonen \etal\ (CDF Coll.), CDF Conf. Note 10761 (2012)

\bibitem{bib:mtophad_cdf}
T. Aaltonen \etal\ (CDF Coll.), FERMILAB-PUB-11-668-E, submitted to Phys.~Rev.~Lett., arXiv:1112.4891 [hep-ex] (2011).

\bibitem{bib:mtopll_d0}
V. M. Abazov \etal\ (D0 Coll.), Fermilab-Pub-12/020-E, submitted to Phys.~Rev.~Lett., arXiv:1201.5172 [hep-ex] (2012).

\bibitem{bib:mtoppole_d0}
V. M. Abazov \etal\ (D0 Coll.), Phys. Lett. B {\bf 703}, 422 (2011).

\bibitem{bib:signlo}
V.~Ahrens \etal, J. High Energy Phys. {\bf1009} (2010) 097, Nucl. Phys. B (Proc. Suppl.) 205-206 (2010) 48.

\bibitem{bib:signnlo}
S.~Moch, P.~Uwer, Phys. Rev. D {\bf78} (2008) 034003; U.~Langenfeld, S.~Moch, P.~Uwer, Phys. Rev. D {\bf80} (2009) 054009.

\bibitem{bib:dm_d0}
V. M. Abazov \etal\ (D0 Coll.), Phys. Rev. D {\bf 84}, 052005 (2011).

\bibitem{bib:dm_cdf}
T. Aaltonen \etal\ (CDF Coll.), CDF Conf.~Note 10777 (2012).

\bibitem{bib:combi}
The Tevatron Electroweak Working Group and CDF and D0 Collaborations, arXiv:1107.5255 [hep-ex] (2011).

\bibitem{bib:mtoptalk}
S. Blyweert, these proceedings.

\end{thebibliography}
\end{document}